\documentclass[12pt]{article}
\usepackage{amssymb,epsf,graphicx}
\newcommand{\be}{\begin{equation}}
\newcommand{\ee}{\end{equation}}
\newcommand{\bea}{\begin{eqnarray}}
\newcommand{\eea}{\end{eqnarray}}
\newcommand{\bdm}{\begin{displaymath}}
\newcommand{\edm}{\end{displaymath}}

\def\<{\langle}
\def\>{\rangle}

\def\a{\alpha}
\def\b{\beta}
\def\c{\gamma}
\def\d{\delta}
\def\g{\gamma}
\def\h{\eta}
\def\lam{\lambda}
\def\m{\mu}
\def\n{\nu}
  
\def\r{\rho}                                     %     \varrho
\def\s{\sigma}                                   %     \varsigma
\def\t{\tau}

\def\D{\Delta}

\def\G{\Gamma}

\def\L{\Lambda}
\def\O{\Omega}

\def\S{\Sigma}

\def\cd{{\cal D}}

\def\cl{{\cal L}}

\def\cn{{\cal N}}
\def\co{{\cal O}}

\def\pa{\partial}
\def\half{{1 \over 2}}

\newcommand{\Tr}{\mathop{\rm Tr}\nolimits}

\newcommand{\im}{\mathop{\rm Im}\nolimits}
\newcommand{\asympt}{\mathop{\sim}}

\def\aver#1{\langle\, #1 \,\rangle}

\def\l{\left}                    % These last three macros I use very often
\def\r{\right}
\let\eps = \varepsilon

\def\LL{{\cal L}}
\def\slashed#1{{\ooalign{\hfil\hfil/\hfil\cr $#1$}}}
\def\L{{\cal L}}
\def\N{{\cal N}}
\def\O{{\cal O}}

\begin{document}
\baselineskip=15.5pt
\pagestyle{plain}
\setcounter{page}{1}
%--------+---------+---------+---------+---------+---------+---------+
%Body

%\begin{document}
\begin{flushright}
CTP-MIT-3514\\
UW/PT 04-05\\
{\tt hep-th/0407073}
\end{flushright}

\vskip 2cm

\begin{center}
{\Large \bf The dual of Janus $\l((<:\l)\leftrightarrow \r(:>)\r)$ an interface CFT \\}
\vskip 1cm

{\bf A.~B.~Clark$^1$, D.~Z.~Freedman$^{2,3}$, A.~Karch$^1$, M.~Schnabl$^3$} \\
\vskip 0.5cm
{\it  $^1$ Department of Physics, University of Washington, \\
Seattle, WA 98195-1560 \\}
{\tt  E-mail: abc@u.washington.edu, karch@phys.washington.edu} \\
\medskip
{\it $^2$  Department of Mathematics, Massachusetts Institute of Technology,\\
Cambridge, MA 02139} \\
{\tt E-mail:dzf@math.mit.edu} \\
\medskip
{\it $^3$ Center for Theoretical Physics, Massachusetts Institute of Technology,\\
Cambridge, MA 02139} \\
{\tt E-mail: schnabl@mit.edu}  \\
\medskip

\end{center}

\vskip1cm

\begin{center}
{\bf Abstract}
\end{center}
\medskip

We  propose and study a specific gauge theory dual of the smooth,
non-supersymmetric (and apparently stable) Janus solution of Type
IIB supergravity found in hep-th/0304129. The dual field theory is
${\cal N}=4$ SYM theory on two half-spaces separated by a planar
interface with different coupling constants in each half-space. We
assume that the position dependent coupling multiplies the operator
$\LL '$ which is the fourth descendent of the primary $\Tr X^{\{I} X^{J\}}$ and
closely related to the ${\cal N}=4$ Lagrangian density. At the
classical level supersymmetry is broken explicitly, but $SO(3,2)$
conformal symmetry is preserved. We use conformal perturbation
theory to study various correlation functions to first and second
order in the discontinuity of $g^2_{YM}$, confirming quantum level
conformal symmetry. Certain quantities such as the vacuum
expectation value $\<\LL'\>$ are protected to all orders in
$g^2_{YM}N$, and we find perfect agreement between the weak coupling
value in the gauge theory and the strong coupling gravity result.
$SO(3,2)$ symmetry requires vanishing vacuum energy,
$\<T_{\m\n}\>=0$, and this is confirmed in first order in the 
discontinuity.

\newpage

\section{Introduction}
\setcounter{equation}{0}

In this paper we describe an application of the
$AdS/CFT$ correspondence to the regular, non-supersymmetric
Janus solution of Type IIB
supergravity found by Bak, Gutperle, and Hirano \cite{janus}.
The solution is the global product of a
5-dimensional domain
wall which approaches $AdS_5$ at the boundary times the
internal space $S^5.$ The novel feature is
that fields are constant along $AdS_4$ slices in the
5-dimensional spacetime, so the isometry group is
$SO(3,2)\otimes SO(6)$.

One might suspect that a non-supersymmetric solution is
unstable, and the stability question was recently examined
in \cite{fakesg}. A generalization of the Witten-Nester
positive energy argument was developed for $AdS_d$-sliced
domain walls.
It was shown that the Janus solution is stable non-perturbatively
within the 5-dimensional theory of the dilaton coupled to gravity.
The proof of stability was extended to include additional fields
which appear in several consistent truncations of Type IIB
supergravity to 5 dimensions. The complete 5-dimensional reduction of
Type IIB supergravity is not known, so the stability of the Janus
solution is not yet fully established. Nevertheless, the present
evidence strongly indicates that the solution is globally stable.

The $AdS/CFT$ correspondence requires that there is a gauge
theory dual for every non-singular solution of Type IIB
supergravity, so it becomes compelling to investigate the
dual field theory for the simple, yet novel, Janus solution.
Indeed the features of this dual were sketched in
\cite{janus}, as we now discuss.

It is well known that the conformal compactification of $AdS_5$ is
contained in the Einstein static universe $R\otimes S^4$. The
compactification includes exactly half the $S^4$ so the boundary
is $R\otimes S^3.$ The conformal compactification of the Janus
solution is also contained in $ESU_5$ as the real line times a
half-space of $S^4$, namely a wedge shaped region with angular
excess. The boundary is thus two hemispheres of $S^3$ joined at a
``corner.'' See Fig. 1. The dilaton field is smooth in the
interior, constant in each hemisphere of the $S^3$ boundary, and
changes discontinuously at the corner. This suggests that the dual
gauge theory is $\cn =4$ super-Yang-Mills theory on $R\otimes S^3$
with a different value of the gauge coupling constant in each
hemisphere. The dual theory would be expected to be conformal,
since $SO(3,2)$ is known to be the conformal group of a defect
conformal theory (dCFT). The non-leading term in the near boundary
asymptotics of the dilaton field suggests a vacuum expectation
value for the operator dual to the dilaton which is identified in
\cite{janus} (and in many other papers) as $\<\Tr F_{\mu\nu}^2\>$.

\begin{figure}[!ht]
\begin{center}
\input{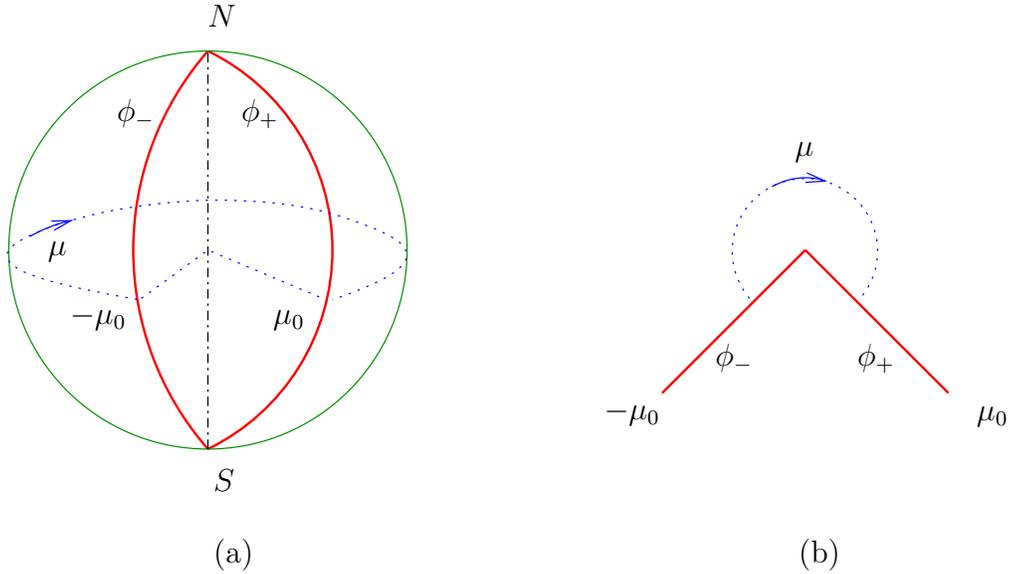}
\caption{\small Conformal spatial geometries of a 2+1 dimensional version of the Janus solution
for (a) global coordinates and (b) for Poincar\'{e} patch coordinates on the $AdS_2$ slices. The boundary
is indicated by bold lines on which the dilaton takes values $\phi_{\pm}$.
In (a) the spatial geometry is the larger ``half'' of a 2-sphere.
In (b) it is the larger ``half'' of the plane, and the boundary is the set of two half lines, which
meet at the defect. The coordinate $z$ in (\ref{mink}) is the radial coordinate in (b), with origin at
the defect.}
\label{Jan-pic}
\end{center}
\end{figure}

The global description of the Janus geometry given above is most transparent if
global coordinates are used for the $AdS_4$ sliced domain wall. If one uses a
Poincar\'{e} patch presentation of the $AdS_4$, one finds that the resulting
5-dimensional coordinate chart includes a boundary region in which two
Minkowski$_{3+1}$ half-spaces are joined at a planar Minkowski$_{2+1}$
interface. This is similar to the defect CFT proposed in \cite{kr1}, \cite{kr2}
and explored in \cite{dfo}. However, there are some important differences which
we discuss below.

We propose and study a detailed version of the gauge theory dual of the Janus
solution. The proposal is based on the assumption that a spatially varying bulk
dilaton is dual to a variant of the Lagrangian density of $\cn =4$ SYM theory
in which the scalar kinetic term is $X^I\Box X^I/2$ rather than the more
conventional $-(\pa X^I)^2/2.$ This operator then multiplies the spatially
varying gauge coupling given by the boundary value of the dilaton. The
discontinuous gauge coupling breaks conformal symmetry from the $SO(4,2)$
bosonic subalgebra of $\cn =4$ SYM theory to the restricted conformal group
$SO(3,2)$ It is an important test of our proposed gauge theory dual that this
symmetry is preserved at the quantum level. At this point we have no general
proof, but we do perform several checks using  conformal perturbation
theory\footnote{Many thanks to Joe Polchinski who suggested this approach.}.

We parameterize the gauge coupling on the two sides of the interface
as $g^{-2}_\pm = g^{-2}_{YM}(1 \mp \gamma)$. The
parameter $\gamma$ governs the magnitude of the discontinuity, and
it is related to the rate of change of the bulk dilaton field
$\phi(\mu)$ with respect to the radial coordinate $\mu$. Our theory
can then be viewed as superconformal $\cn =4$ SYM theory perturbed
by the operator $\gamma \, \eps(z_3) \, \LL'(z)$, where $\LL'(z)$ is
the variant of the $\cn =4$ SYM  Lagrangian described above,
$\eps(z_3)$ is the step function with values $\pm 1$, and $z_3$ is
the Cartesian coordinate perpendicular to the interface. Since
$\LL'$ is the fourth SUSY descendent of the primary $\Tr X^{\{I} X^{J\}}$,
it is a
marginal operator. The perturbation breaks supersymmetry because of
the step function.

In conformal perturbation theory effects of the interface on
correlation functions can be computed order by order in $\g$
but to all orders in $g^2_{YM}N$. For example, the vacuum expectation
value $\< \LL'\>$ vanishes in the
unperturbed theory. Conformal perturbation theory implies that
 $\< \LL'\>$ can be expressed as the series
\be \label{cpt}
\< \cl'(z) \>_{\rm Janus}  =  - \gamma \int d^4x \eps(x_3) \<\cl'(z)
\cl'(x) \> + \frac{\gamma^2}{2} \int d^4x d^4y\, \eps(x_3) \eps(y_3)
\<\cl'(z) \cl'(x) \cl'(y) \>  + \co(\gamma^3)
\ee
The 2- and 3-point correlation functions which appear in the integrals are {\it
protected} correlators of the {\it unperturbed} $\cn = 4$ theory. They are
independent of $g^2_{YM}$ to all finite orders \cite{minwal}, \cite{dhfs}, \cite{Howe},
\cite{intril}, \cite{Basu} and have no instanton corrections \cite{bianchi}. Therefore,
through order $\g^2$, the VEV $\< \LL'(x)\>$ should be protected and have the
same value whether calculated at weak coupling in the gauge theory or from the
bulk solution at strong coupling. It turns out that the three-point function in
(\ref{cpt}) vanishes due to supersymmetry in the unperturbed SYM theory, and
the order $\g$ term involving the two-point function can be easily evaluated
and agrees perfectly with the value obtained from the bulk Janus solution!

More generally, the order $\g$ correction to any correlation
function $\< \co_1(x_1) \, \ldots \co_n(x_n) \>$ is given by
\be
\< \co_1(x_1) \, \ldots \co_n(x_n) \>_{\g}
= - \gamma \int d^4z \; \eps(z_3) \< \co_1(x_1) \, \ldots \co_n(x_n)
\cl'(z)\>_{\cn=4}.
\ee
We calculate several examples of 2-point functions of gauge-invariant operators
and show that their order $\g$ contributions are compatible with defect
conformal symmetry \cite{Cardy}, \cite{McAO}.

The present theory may be compared with the earlier dCFT of \cite{kr2},
\cite{dfo} in which a probe $D5$-brane wrapped on an $S^2 \subset S^5$ has an
equilibrium position on an $AdS_4 \subset AdS_5$. In the dual gauge theory the
fields of $\cn =4$ SYM interact with $D5$-brane modes confined to the defect.
The $R$-symmetry is $SO(4)$  and the preserved superalgebra is $OSp(4,4)$ which
contains $SO(3,2)\otimes SO(4)$ as the maximal Lie subalgebra. The Janus
solution maintains $SO(6)~R$-symmetry, but breaks supersymmetry completely.
There is no brane interpretation of the defect, so we conclude that the dual
gauge theory contains no degrees of freedom on the defect. Instead there is a
discontinuous gauge coupling which simulates a dielectric interface. For these
reasons we prefer the name interface conformal field theory (ICFT).
A different class of dilatonic deformations of $AdS_5 \otimes S^5$ 
was studied recently
in \cite{ohta}. Their solutions break $SO(3,2)$ and develop thus nontrivial
VEV for $\< T_{\m \n} \>$.  
It might be interesting to study the gauge theory dual for
their deformations using conformal perturbation theory.

In Sec. 2 below we review the Janus solution of Type IIB supergravity
giving new details on the boundary behavior of the dilaton and metric.
In Sec. 3 we discuss the computation of the vacuum expectation values
 $\< \LL'\>$ and $\<T_{\m\n}\>$ from the gravity theory.  In Sec 4. we
present our first proposal for the gauge theory dual via conformal perturbation
theory. In Sec. 5 we present calculations of various one- and two-point
functions, and compare with gravity. In Sec. 6, we outline another approach to
the Janus dual in which we specify a Lagrangian in which the fields obey
specific jump conditions at the interface. In an Appendix A we consider general
$\cn =1$ SUSY theories with a position dependent coupling. We show that two of
the four $\cn =1$ supercharges are conserved if certain localized interface
``counterterms'' are added to the Lagrangian. In the $\cn =1$ form of the $\cn
=4$ SYM, the counterterms break $SO(6)$ and are therefore not permitted in the
Janus dual, which therefore breaks all supersymmetry explicitly.

{\bf Note added:} We thank K. Skenderis for informing us that his paper with
I.~Papadimitriou entitled ``Correlation function in holographic RG flows'' will soon be submitted
to hep-th. This paper includes a discussion of holography for the Janus solution.

\section{The Janus solution.}
\setcounter{equation}{0}

The Janus solution involves the metric, dilaton, and 5-form
of Type IIB supergravity
\footnote{The relevant 10-dimensional equations of motion
are ~~$R_{MN} =
\half\pa_M\phi\pa_N\phi+\frac{1}{96}F_{M....}F_N^{....};
\newline~~\Box_{10}\phi=0;~~*F_5=F_5.$}
with other fields vanishing:

\bea \label{jan}
ds_{10}^2 &=& L^2e^{2A(\mu)} \l( g_{ij} dx^idx^j + d\mu^2 \r)
+ L^2 ds^2_{S^5}\\
\phi &=& \phi(\mu)\\
F_5 &=& 4L^4 \l[e^{5A(\mu)}d\mu\wedge\omega_{AdS_4}\,+\,\omega_{S^5} \r]\\
   &=& dC_4\,+\,*dC_4\\
C_4&=& h(\mu) \omega_{AdS_4}\\
h'(\m) &=& 4L^4 e^{5A(\mu)}\,,
\eea
where $g_{ij}$ is an $AdS_4$ metric of unit scale, and
$ds^2_{S^5}$ is a metric on the unit $S^5$. The volume forms
are also normalized to unit scale. We present the  $AdS_4$
metric in both global and patch coordinates, viz.
\be \label{glo}
ds_4^2 = \frac{1}{\cos^2\lam} \l[-d\t^2 + d\lam^2 + \sin^2\lam \, d\Omega_2^2 \r],
\ee
with $0 \le \lam < \half \pi$ and $ d\Omega_2^2$ a metric on the
unit 2-sphere, or
\be \label{patch}
ds_4^2 = \frac{1}{z^2}\l[-dt^2 + dx^2 +dy^2 + dz^2\r]\,.
\ee
where $z \ge 0$ and $-\infty < t,x,y < \infty$.
% {\bf Please check all of (\ref{jan}).}

The equations of motion of constrain
the dilaton and scale factor to obey (\cite{janus},
\cite{fakesg})
\bea \label{eom}
\phi'(\m) &=& c e^{-3A(\mu)}\\
A'(\m) &=&\sqrt{e^{2A} -1 +be^{-6A}}\,.
\eea
The arbitrary parameter\footnote{The dimensionless parameter
 $c$ in (\ref{eom}) is related to the $c$ used in
\cite{fakesg} by a factor of $L$.}   $c$ governs the rate of
 variation of the dilaton, and $b=\frac{c^2L^2}{24}$.
The equation for $A'(\m)$ cannot be integrated in closed form, and
one must be content with the implicit solution \be \label{imp} \m
= \int_{A_0}^{A}\frac{dA}{\sqrt{e^{2A} -1 +be^{-6A}}}\,. \ee The
parameter $A_0=-\ln(x_{min})/2$ where $x_{min}$ is the
smallest root of the rational function $P(x) \equiv 1-bx^3-x^{-1}$
which appears in the denominator of (\ref{imp}) with argument
$x=e^{-2A}.$ The resulting geometry is free of curvature
singularities as long as the two roots of $P(x)$ remain distinct,
namely for $0<c<\frac{9}{4\sqrt{2} L}$.

%{\bf We now work with $\k^2/2=1$. }
%\noindent

The boundary of the 5-dimensional space-time is reached at the value
of the radial coordinate $\m_0$, which can be expressed as the
series \cite{fakesg} \be \label{bdy} \m_0 =
\frac{\sqrt{\pi}}{2}\sum_{n=0}^{\infty}\frac{\G(4n+\half)}{\G(3n+1)}\frac{b^n}{n!}\,.
\ee Near the boundary, we have the behavior \be \label{bdybeh}
e^{A(\m)} \sim \frac{1}{\sin(\m_0-\m)}[1 + O(b \sin^8(\m_0-\m))]\,,
\ee which shows that the back reaction of the flowing dilaton is of
order $(\m_0-\m)^8$ near the boundary.

Actually the equations above define the geometry only in the
region $0<\m< \m_o.$ However one can smoothly continue
$A(\m)$ as an even function to the region
$-\m_0 < \m<\m_0$. This is the full geometry which is
geodesically complete with two boundary regions, namely
$\m \to \pm \m_o.$

The dilaton equation in (\ref{eom}) is formally integrated to give
\bea \label{dil}
\phi(\m) &=& \phi_0 + c\int_0^\m e^{-3A(\m)}
d\m\\
&=& \phi_+ - c\int_\m^{\m_0} e^{-3A(\m)} d\m\,. \eea Note that the
deviation from the (arbitrary constant) value $\phi_0$ has the
opposite sign in the two regions $\rm{sgn(\m)} =\pm$. The dilaton
approaches the boundary values \be \label{dilbo} \phi_{\pm} =
\phi_0 \pm c \int_0^{\m_0} e^{-3A(\m)}d\m\,. \ee One can derive
the series representation
\be \label{dilser} \phi_{\pm} =\phi_0
\pm \frac{c\sqrt{\pi}}{2}\sum_{n=0}^\infty
\frac{\G(4n+2)}{\G(3n+\frac{5}{2})}\frac{b^n}{n!}\, = \phi_0 \pm
\frac{2}{3} c + \co(c^3). \ee
From this relation we see that the
parameter $\gamma$ we introduced in the introduction is determined
in terms of the supergravity solution as
\be \label{gammac} \gamma
= \tanh\l(\frac{\phi_+ - \phi_-}{2}\r) = \frac{2}{3} c +
     \frac{16}{189} c^3 + \co(c^5) \ee
We also need the dominant subleading term which can be obtained by
inserting (\ref{bdybeh}) in the second integral of (\ref{dil}).
The result may be expressed as \be \label{dilasym} \phi(\m)
\asympt_{\m \to \pm\m_0} \phi_\pm \mp \frac{c}{4} e^{-4A(\m)} +
O\l(e^{-6 A(\m)}\r)\,. \ee

Let us now discuss the issue of the boundary metric. We must keep
in mind the fact that the interior metric diverges on the boundary
and thus determines the conformal structure there but not quite
the metric \cite{witten}. To obtain a specific boundary metric one
must multiply the 5-dimensional space-time metric of (\ref{jan})
by a conformal factor $f^2$ where $f$ has a linear zero at the
boundary. If we choose $f=e^{-A(\m)}$, for example, then the
boundary appears as two copies of $AdS_4$, with global metric,
joined at their common boundary $R\otimes S^2.$

The determination of the boundary metric is important for the
AdS/CFT correspondence because the gauge theory dual is coupled to
the specific metric obtained. We will need to discuss several
choices in this paper, and the first is the case of two copies of
$AdS_4$ discussed above \cite{porrati}. Another case is the conformal
compactification of the Janus solution, in which one uses the
global $AdS_4$ metric on slices, and takes $f =
e^{-A(\m)}/\cos\lambda$. As discussed in \cite{janus} and
\cite{fakesg}, the 5-dimensional Janus spacetime metric is then
written as
\be \label{cc}
ds^2_5 = \frac{L^2
e^{2A(\m)}}{\cos^2\lambda}\l[-dt^2  + d\lambda^2
+\sin^2\lambda\, d\Omega_2^2 +\cos^2\lambda \, d\m^2\r]\,.
\ee
The compactified 5-metric in (\ref{cc}) is then the portion of
$ESU_5$ containing half of $S^4$ with angular excess (since
$\m_0>\frac{\pi}{2}$) and thus two corners at the poles. See Fig.
1(a). The boundary consists of the two regions obtained in the limits
$\mu \to \pm \m_0.$ Each spatial region is a hemisphere of $S^3$,
and the two hemispheres are joined at the poles of the $S^4$. The
boundary metric is thus $R\otimes S^3$, as in the global boundary
of the $AdS_5\otimes S^5$ solution of IIB supergravity. However,
in the Janus solution the dilaton is not constant on the full boundary,
but rather takes the values in (\ref{dilbo}).

The final choice involves the patch metric (\ref{patch}) on the
$AdS_4$ slices and the scaling factor $f=e^{-A(\m)}z$. The Janus
metric is then
\be \label{mink}
ds_5^2 = f^{-2}\l[ -dt^2\,+\,dx^2\,+\,dy^2\,+\,dz^2\,+\,z^2d\m^2\r]\,,
\ee
and the conformal metric in (\ref{mink}) is the product of Minkowski$_{2+1}$
and a wedge (with radial coordinate $z$ and angular range $-\m_0<\m<\m_0$). The
boundary is then two copies of Minkowski$_{3+1}$ joined at the
Minkowski$_{2+1}$ interface located at $z=0$. The dilaton takes the values
$\phi_{\pm}$ in (\ref{dilbo}) on the half-spaces at $\mu=\pm \mu_0$. See Fig. 1(b).

\section{Gravity predictions for the dual gauge theory.}
\setcounter{equation}{0}

The AdS/CFT correspondence permits the calculation, from the gravity
side of the duality, of correlation functions of the CFT coupled to
the specific boundary metric obtained by the procedure discussed in
Sec. 2. Namely, one cancels the singularity of the bulk space-time
metric by multiplication by $f^2$, where $f$ has a linear zero at
the boundary. Different choices of $f$ give different boundary
metrics, and correlation functions of operators ${\cal O}_\D$ for a
pair of boundary metrics $g_{(1)}$ and $g_{(2)}$ are related by
multiplication by factors $(f_{(1)}/f_{(2)})^\D$ for each operator.
This is required by conformal symmetry.  For each choice of $f$, the
behavior near the boundary of the dual bulk field $\phi_\D$ can be
expressed in terms of $f$ and boundary coordinates $x^\m$ as
\be \label{andreas}
\phi_\D \sim a_f(x) \l(f^{d-\D}+\ldots\r) + b_f(x) \l(f^\D + \ldots\r).
\ee
The coefficient $a_f(x)$ of the leading term is the source (or coupling) of the
operator in the deformed CFT, while the vacuum expectation value \cite{vj1},
\cite{vj2} is given by $\<{\cal O}_\D(x)\> = -(2\D-4) b_f(x)$. Since $\phi_\D$
is a scalar, the products $a_f(x)\,f^{d-\D}$ and $b_f(x)\,f^\D$ are invariant
under change of boundary metric. Couplings and vacuum expectation values thus
change correctly.

Suppose, for example, that we have a bulk scalar with asymptotic
behavior $a_{\pm} \, e^{(d-\Delta) A(\mu)}$, corresponding to a
piecewise constant coupling $a_{\pm}$ for the dual operator ${\cal
O}_\D$ on the two $AdS_4$ halves of the boundary. Then there is a
position dependent coupling $\frac{a_{\pm}}{z^{d-\Delta}}$ for the
Minkowski space ICFT. For the special case of the dilaton coupling
to the $\Delta=d$ Lagrangian density operator, a piecewise constant
coupling in AdS remains piecewise constant in Minkowski space.
Similarly, a constant one point function
\be \label{adsvev}
\< {\cal O} \>_{AdS} = v\ee
for an operator on two copies of $AdS_4$
becomes
\be \label{minvev}
\< {\cal O} \>_{M} = \frac{v}{z^{\Delta}}
\ee
for the same operator in the
Minkowski space ICFT. 
As noted in \cite{adfk} this reproduces correctly the form of the
one-point function required by defect conformal symmetry.

We are mostly interested in the bulk dilaton $e^{\phi(\m)}$ of the Janus
solution (\ref{jan}). We will argue in Sec. 4 that it is dual to an operator
${\cal L}'$ which is closely related to the Lagrangian of ${\cal N}=4$ SYM
theory and is a marginal operator with $\D=d$. The coupling and VEV of ${\cal
L}'$ for different boundary metrics can be read from the AdS/CFT formula
 (see \cite{vj2}, \cite{magoo})
\be\label{magoovev}
e^{\phi(\m)} = e^{\phi_\pm} \l( 1- \frac{2\pi^2}{N^2} \aver{\cl'}
f^4 + \ldots \r).
\ee
The leading term gives the gauge coupling in the dual ICFT
which is the same for all
boundary metrics with the same conformal structure, whether two
copies of $AdS_4$ or two half Minkowski spaces, and is given by
\be\label{gcoup}
e^{\phi_\pm}= g_s = \frac{g_{YM\pm}^2}{2\pi},
\ee
with $\phi_\pm$ in (\ref{dilser}). The subleading term determines the VEV which
is obtained by comparison with (\ref{dilasym}). For $f=e^{-A}$ which yields two
copies of $AdS_4$ as the boundary metric, we find
\be \< {\cal L}' \>_{AdS} = \eps(z)
\frac{N^2}{2 \pi^2} \frac{c}{4}\,.
\ee
For $f=z
e^{-A}$, which yields the Minkowski space ICFT, the result is
\be\label{lgrav}
\< {\cal L}' \>_{M} = \eps(z) \frac{N^2}{2 \pi^2}
\frac{c}{4 z^4} \,.
\ee
This result obtained from gravity will be compared with gauge theory
in Sec. 5.

The Janus metric in (\ref{jan}) also contains holographic information about the
stress tensor of the boundary ICFT. Although non-vanishing vacuum expectation
values for scalar operators are expected in an ICFT, the VEV of the stress
tensor $T_{\m\n}(x)$ must vanish \cite{McAO}. To
see that AdS/CFT gives the required result, we must examine the Janus metric
with the various conformal factors discussed above. For example, with
$f=ze^{-A}$, (\ref{mink}) shows that the transverse metric is $\h_{\m\n}$ for
all values of the radial coordinate $\m$. Hence the traceless component
required by defect conformal symmetry 
vanishes, and we find the result $\<T_{\m\n}\>=0$ as an exact prediction of
supergravity.

It is also interesting to consider the correlation functions $\aver{\Tr
X^{k_1}\, \Tr X^{k_2}\dots \Tr X^{k_n}}$ of chiral primary operators in the
Janus geometry. In supergravity one would calculate these correlators from
Witten diagrams involving bulk-to-boundary propagators and vertices, However,
it is easy to see that, to order $c$, the Janus solution has no effect, since
the dilaton appears in the Einstein frame action of Type IIB supergravity only
in the terms
\be \label{sgact}
\cl = \half \sqrt{-G} \l[ \pa_\m \phi \pa^\m \phi + e^{-2\phi} \pa_\m
A  \pa^\m A \r]
\ee
(where $A$ is the axion). Thus effects on the dynamics of the bulk fields
$h_{\a\b}$ and $a_{\a\b\c\d}$, which are dual to the SYM operators $\Tr X^k$,
begin in order $c^2$.

%%%%%%%%%%%%%%%%%%%%%%%%%%%%%%%%%%
%%%%%%%%%%%%%%%%%%%%%%%%%%%%%%%%%
\section{Proposed gauge theory dual}
\setcounter{equation}{0}

In this section we will outline the first of two closely related
proposals for the gauge theory dual of the Janus solution. We
emphasize that they are proposals rather than a derivation. The two
proposals produce the same results for several tests discussed in
Sec. 5 in which we work to order $\g$ and $\g^2$ in the
discontinuity of the gauge coupling across the interface.

The main guidepost for the initial proposal is the hypothesis that a spatially
varying dilaton couples to the marginal operator which is the fourth
supersymmetric descendent of the chiral primary $\Tr X^{\{I}X^{J\}}$. This
descendent is closely related to the usual Lagrangian density of ${\cal N}=4$
SYM theory, differing by a total derivative which changes the scalar kinetic
term from $-\half \Tr D_\m X^I D^\m X^I$ to $\half \Tr X^I D^\m D_\m X^I$,
where $D_\m X^I =\pa_\m X^I + i[A_\m,X^I]$. We call this operator ${\cal
L}'_0$. It contains all interactions of  ${\cal N}=4$ SYM with no explicit
gauge coupling.

We therefore hypothesize that the gauge theory action with
spatially varying coupling $g^2(x)$ takes the form
\be \label{genact}
S = \int d^4x \frac{1}{g^2(x)} {\cal L}_0'(x).
\ee
We specialize to the case of the Janus dual where the coupling
is constant on the two sides of the interface located at
$x^3=0$, so that
\bea \label{coup}
\frac{1}{g^2(x)} &=& \frac{1}{g^2_+}\theta(x^3) +
\frac{1}{g^2_-}\theta(-x^3)\\
&=& \frac{1}{\bar{g}^2}\l[1 - \g \eps(x^3)\r],
%&=& \frac{1}{\bar{g}^2}\l[ \frac{1}{1-\g^2} +
%\frac{\g}{1-\g^2}\eps(x^3)\r]
\eea
in which $\bar{g}^{-2} =\frac{1}{2}\l(g_+^{-2} + g_-^{-2}\r)$ is the
average coupling, and
\be \label{gam}
\g = \frac{g_+^2 - g_-^2}{g_+^2 + g_-^2}
\ee
is a measure of the jump in the coupling across the interface.
(Note that $\theta(x^3)$ and $\eps(x^3)$ are standard step
functions.)
%We then define new average coupling
%$\hat{g}^2 = \frac{2}{g_+^{-2} + g_-^{-2}} = \bar{g}^2(1-\gamma^2)$ and
We then write our proposal for the Janus dual action as
\bea \label{janact}
S   &=& S_0 + S'\\
S_0 &=& \int d^4x {\cal L}'(x) \\
S'  &=&  -\gamma \int d^4x \eps(x^3) {\cal L}'(x),
\eea
where $\L'=\frac{1}{\bar{g}^2} \L_{0}'$.

If this proposal is correct for the dual to Janus then it must
yield a quantum theory with broken supersymmetry and $SO(3,2)$
defect conformal symmetry. Let us first discuss these properties
at the level of the classical action (\ref{janact}).

It is well known that the key to conventional $SO(4,2)$ conformal
symmetry is proper behavior under the conformal inversion $x'^\m
= \frac{x^\m}{x^2}$, since finite special conformal
transformations are composed from inversion and translations.
Inversion is also a symmetry of ICFT, since it transforms the
defect at $x^3=0$ into itself. Special conformal transformations
in the preserved conformal group $SO(3,2)$ are composed from
inversion and translations in the $x^{0,1,2}$ directions.
In conventional ${\cal N}=4$ SYM theory, the
operator ${\cal L}'(x)$ transforms as a field of dimension 4,
namely ${\cal L}'(x) = (x')^8 {\cal L}'(x')$, and the volume
element transforms as $d^4x = d^4x'/(x')^8.$  Clearly
$\eps(x^3) = \eps(x'^3)$, so $S'$ is inversion symmetric and
thus $SO(3,2)$ invariant at the classical level.

Although $S_0$ in (\ref{janact}) enjoys full ${\cal N}=4$
supersymmetry, it is fairly easy to see that supersymmetry is
explicitly and completely broken in $S'$ because of the step
function. The straightforward technical argument for this is
presented in Appendix A.

In conventional ${\cal N}=4$ SYM theory, with constant gauge
coupling, one can integrate by parts in the action, and it is
immaterial whether one writes the Lagrangian density as the
conventional ${\cal L}$ or as  ${\cal L}'$. Integration by parts
cannot be used in $S'$, so the choice of Lagrangian density is
significant. With the choice ${\cal L}'$, which is a marginal
operator in the undeformed theory, one may hope that conformal
symmetry holds in our proposal for an ICFT.
Since ${\cal L}' - {\cal L} = \frac{1}{4}\Box {\rm Tr}(X^IX^I)$,
 which is the
total derivative of the Konishi scalar with positive anomalous
dimension, there would be no chance of conformal symmetry at the
quantum level if one replaced ${\cal L}'$ by ${\cal L}$ in $S'$.

We have seen that the proposed action for the
Janus dual has the correct properties at the classical level.
Our goal now is to explore the proposal at the quantum level,
specifically to undertake calculations to test $SO(3,2)$
conformal symmetry and match quantities which can be computed
in the bulk theory.

The most interesting and potentially powerful way to proceed is to use
conformal perturbation theory. Correlation functions are defined
through the formal path integral
\be \label{cpt1}
\<\co_1(x_1) \, \ldots \co_n(x_n) \> = \int \cd \Phi  \;
e^{S_0+S'} \, \co_1(x_1)\,\ldots \, \co_n(x_n)\,,
\ee
in which $S'$ is viewed as a perturbation. One can expand
in powers of $\gamma$ and obtain an expression for all $n$-point
functions in the perturbed theory in terms of correlation
functions in the undeformed ${\cal N}=4$ SYM
\bea \label{cpt2}
\< \co_1(x_1) \, \ldots \co_n(x_n) \>_{\rm Janus}
&=& \< \co_1(x_1) \, \ldots \co_n(x_n) \>_{\cn=4}\nonumber\\
&& - \gamma \int d^4z \; \eps(z_3) \< \co_1(x_1) \, \ldots
\co_n(x_n) \cl'(z)\>_{\cn=4}
\nonumber\\
&& + \frac{\gamma^2}{2} \int\!\!\int d^4z\, d^4w \, \eps(z_3) \eps(w_3)
\, \< \co_1(x_1) \, \ldots \co_n(x_n) \cl'(z) \cl'(w)\>_{\cn=4}\nonumber\\
&& + O(\gamma^3)\,.
\eea
In a number of interesting cases the ${\cal N}=4$ correlation
functions which appear in (\ref{cpt2}) are known exactly. Then we
can obtain results for the effects of the discontinuous coupling
to fixed order in $\gamma$ but all orders in the 't Hooft coupling
$g^2_{YM}N$\,! It is exactly for these ``protected" quantities
that results from gauge theory theory should match results from
gravity!

\section{Calculations using conformal perturbation theory}
\setcounter{equation}{0}

In this section we describe several calculations to order $\g$ and $\g^2$ of
correlation functions in the Janus dual field theory. Our goals are two-fold,
first to test $SO(3,2)$ conformal symmetry at the quantum level, and second to
match results with those obtained from the gravity side of the duality. The
$\cn =4$ correlators which appear in the CPT integrals are $SO(4,2)$ covariant,
so the integrals are formally invariant under the $SO(3,2)$ subgroup that
preserves the defect at $z_3=0$. However the integrands are singular when an
integrated point, e.g. z approaches one of the $x_i$. Thus we need to regulate
the integrals and study them before $SO(3,2)$ invariance can be claimed.  The
main regularization method used involves a spatial cutoff, that is we exclude a
small interval $|z_3 - x_3| \,<\, \delta$, integrate and then take the limit as
$\delta \to 0$. With this prescription contact terms in the $\cn =4$ correlators
do not contribute. Any {\it protected} $\cn =4$ correlator can be evaluated in
the free field limit in which the terms $\half(X\Box X +
\bar{\lam}\slashed{\pa}\lam)$ in $\cl'$ {\it only} contribute such contact
terms. Thus we can restrict to $\cl' \to -\frac{1}{4g^2_{YM}} F^2$ for
protected correlators.

\subsection{$1$-point function of $\cl'$}
\setcounter{equation}{0}
% and $T_{\m\n}$}

The first application of conformal perturbation
theory is the one-point function of $\cl'$. Using (\ref{cpt2}) we
have
\be \label{lvev}
\< \cl'(x) \>_{\rm Janus}  =  - \gamma \int d^4z \eps(z_3) \<\cl'(x)
\cl'(z) \> + \frac{\gamma^2}{2} \int d^4z d^4w\, \eps(z_3) \eps(w_3)
\<\cl'(x) \cl'(z) \cl'(w) \>  + \co(\gamma^3) \ee
All terms on the
right hand side are correlators in the unperturbed $\cn=4$ theory
and we have used the fact that all one-point functions vanish in a
CFT on undeformed Minkowski spacetime.

The order $\gamma$ term involves the two-point function $\aver{\cl'(x)
\cl'(z)}$. This is a protected quantity which can be evaluated in the free
$\cn=4$ theory, and the result will hold to all orders in $g^2_{YM}N$. In the
free theory it is straightforward to compute
\be \label{lcorr}
\aver{\cl'(x) \cl'(z)}_{\cn=4} = \frac{3}{\pi^4}
\frac{N^2}{(x-z)^8}~~~~~~~~~~z \ne x\,,
\ee
where the result comes from the  kinetic term for the gauge field.
(Had we used $\cl$ instead of $\cl'$ the scalars would contribute
by changing $3 \to 3 + 6 \times \frac{3}{2} = 12$.) The order
$\g$ term in (\ref{lvev}) involves a singular integral which we
perform using a cutoff:
\bea \label{inte}
I &=&  \int  d^4z \, \eps(z_3) \frac{1}{(x-z)^8} =
\int dz_3 \, \eps(z_3) \int d\vec{z} \frac{1}{\l((x_3-z_3)^2+{\vec{z}}^2\r)^4}
\nonumber\\
&=& \frac{\pi^2}{8} \int_{|z_3-x_3|>\delta} \!\!\! dz_3 \, \frac{\eps(z_3)}{\l|x_3-z_3\r|^5}
=  \frac{\pi^2}{8} \l(\frac{1}{2\delta^4} -\frac{1}{2x_3^4}\r) \eps(x_3)\,.
\eea
The quartic divergent term can be cancelled by adding an additive
constant to $\cl'$ in $S'$ of (\ref{janact}). Contact terms from
$\half(X\Box X + \bar{\lam}\slashed{\pa}\lam)$ which were omitted
in  (\ref{lcorr}) give similar divergent contributions.
The remaining term has the $1/(x^3)^4$ dependence required by
defect conformal
symmetry. Inserting it in (\ref{lvev}) yields the result\footnote{The same
finite result (with no divergent term) can be obtained using differential
regularization \cite{diff}. One simply uses the identity $\frac{1}{(x-z)^8} =
\frac{1}{24}\Box_z\frac{1}{(x-z)^6}$ in (\ref{inte}), and integrates by parts.
After use of $(\pa_{z_3})^2 \eps(z_3) = 2 \d'(z_3)$, one finds a convergent
integral over $\vec{z}$ which is easily performed.}
\be \label{triumpf}
 \< \cl' \>_{{\rm Janus}} = \eps(z_3) \frac{3N^2}{16 \pi^2} \frac{\gamma}{z_3^4} + \co(\g^3).
\ee

We have written the remainder as $\co(\g^3)$, because the integrand of the
order $\g^2$ term in (\ref{lvev}) involves the 3-point correlator
$\aver{\cl'(x)\cl'(z) \cl'(w)}$ which actually vanishes. This is a 3-point
function of descendants of the chiral primary $\Tr X^{\{I} X^{J\}}$. It is
protected, and we again have the luxury of working at the free field level at
which we can restrict our attention to the gauge field terms: $\aver{F^2(x)
F^2(y) F^2(z)}$. To show that this 3-point function vanishes, just express
$F^2$ in terms of self-dual and anti-self-dual parts of field strengths,
\bea
(F_{\m\n})^2 &=& (F^{+}_{\m\n})^2 +(F^{-}_{\m\n})^2\\
\nonumber F^{\pm}_{\m\n} &=& \frac{1}{2} ( F \pm \tilde{F})_{\m\n}\,.
\eea
Using the direct terms in (\ref{fdir}), one can easily verify that
\be\label{Wick}
\<F^+_{\m\n}(x) F^+_{\rho\s}(y) \> = \<F^-(x)_{\m\n} F^-(y)_{\rho\s} \>=0\,.
\ee Hence only $\<F^+_{\m\n}(x) F^-_{\rho\s}(y)\>$ contractions contribute. For
a $k$-point function we have $k$ contractions. Since each of them contracts one
$F^+$ and one $F^-$ we need an equal number $k$ of each. Since the number of
$F^+$ is even, $k$ must be even. If $k$ is odd the $k$-point function of $F^2$
automatically vanishes. This applies to the tree approximation in any gauge
field theory and to all orders for $\aver{\cl'(x)\cl'(z) \cl'(w)}_{\cn=4}$
because it is a protected correlator.

The result (\ref{triumpf}) for $\aver{\cl'}$ is in perfect agreement
with the gravity result (\ref{lgrav}).  In Sec 6.1 we show that one
obtains the same result in the weak coupling limit using
perturbation theory with the propagator (\ref{bothpos}). It is
conformal perturbation theory with protected correlators which
allows us to understand this agreement between field theory and
gravity, which is a strong test for the proposed Janus dual and a
precise new test of the AdS/CFT correspondence. Since $n$-point
functions in the $\cn=4$ theory are generically not protected for
$n\ge4$, we do not expect agreement to hold to all orders in
$\gamma$.

\subsection{$\aver{T_{\m\n}(z)}_{{\rm Janus}}$}
%$\aver{T_{\m\n}(z) \cl'(x)}_{{\rm Janus}}$ }

The one-point function of the stress tensor must vanish in a field theory
with $SO(3,2)$ defect conformal symmetry  \cite{McAO}, and we have already
seen that it is does vanish when calculated from the bulk Janus solution.
It is therefore an important test to check whether it vanishes
in our proposed dual gauge theory.

In conformal perturbation theory the 1-point function is given by
\bea\label{Tcpt}
\aver{T_{\mu\nu}(z)}_{\rm Janus} &=&  \aver{T_{\mu\nu}(z)}_{\N=4}
-\gamma \int d^4x \, \eps(x_3) \aver{T_{\mu\nu}(z) \L'(x)}_{\N=4}
\nonumber\\
&& + \frac{\gamma^2}{2}  \int d^4x d^4y \, \eps(x_3) \eps(y_3)
\aver{T_{\mu\nu}(z) \L'(x) \L'(y)}_{\N=4} + O(\gamma^3)
\eea
The one point function on the right hand side vanishes trivially in
the undeformed $\cn =4$ theory, and we will now argue that the order
$\g$ contribution also vanishes.

There is an interesting $\cn =4$ SUSY argument to show that
$\aver{T_{\m\n}(z) \cl' (x)}_{\cn = 4}$ vanishes. For any pair of
operators $F(x),\,B(y)$, the first fermionic and the second bosonic,
there is a basic identity ((see (8.1) of \cite{intril})
\be
\aver{\{Q,F(x)\}\,B(y)}=\, \aver{F(x)\,[Q,B(y)]},
\ee
valid if the vacuum is supersymmetric (i.e. $Q|0\>=0$) In our case both
operators are descendants of the primary ${\cal O}_{20'}=\Tr X^{\{I} X^{J\}}$
and we can write (with indices and brackets $[...]_{\pm}$ suppressed)
\bea \label{susy}
\aver{T(z) \cl'(x)} &\sim& \aver{Q^2\bar{Q}^2{\cal O}_{20'}(z)(Q^4
+\bar{Q}^4){\cal O}_{20'}(x)}\\ \nonumber &=&
\aver{Q\bar{Q}^2{\cal O}_{20'}(z) Q^5{\cal O}_{20'}(x)} +
\aver{\bar{Q}Q^2{\cal O}_{20'}(z) \bar{Q}^5{\cal O}_{20'}(x)}~=~0\,.
\eea
In the last line we have used the special BPS condition $Q^5{\cal O}_{20'}=0$
satisfied by the lowest dimension chiral primary (see (4.3) of \cite{intril}).
The ordering of operators in $Q^2\bar{Q}^2{\cal O}_{20'}$ does not affect this
argument, since reordering produces $P^\m$ descendants, which have vanishing
2-point functions in (\ref{susy}).

An alternative way to prove that $\aver{T_{\mu\nu}(z) \L'(x)}=0$ is
to use the known nonrenormalization theorem. A short glance at
the free field contribution shows that scalar and fermion
contributions vanish by their equations of motion. The
gauge field contribution vanishes because
\bea
\L'_{gauge} &=& -\frac{1}{4g^2} \l[ \l(F_{\mu\nu}^+\r)^2 +
\l(F_{\mu\nu}^-\r)^2 \r]
\\
T_{\mu\nu}^{gauge} &=& \frac{1}{g^2} \l( F_{\mu\rho}^+
F_{\nu}^{-\,\rho} + F_{\mu\rho}^- F_{\nu}^{+\,\rho} \r)
\eea
and therefore the free contractions vanish by (\ref{Wick}).

In order $\g^2$ we encounter something new. The correlator
 $\aver{T_{\m\n} \cl' \cl'}_{\cn =4}$ is protected and does {\it not}
vanish. It cannot vanish because the translation Ward identity relates it to
$\aver{\cl'\,\cl'}_{\cn =4}$. The tensorial form of this 3-point function is
known \cite{Petkou}, \cite{McAO}, \cite{liu} but complicated, and we postpone 
the study of a second order contribution to $\aver{T_{\mu\nu}}$ to the future.    

\subsection{Two point functions}

Let us now study more generally whether our theory based on
conformal perturbation theory with perturbing operator $\gamma
\eps(z_3) \cl'(z)$ does indeed lead to $SO(3,2)$ invariant interface
CFT.
In this subsection we will show that the two-point function
\be
\aver{\O(x) \O(y)}_{\rm Janus} = \aver{\O(x) \O(y)}_{\N=4} -\gamma \int
d^4z \, \eps(z_3) \aver{\O(x) \O(y) \L'(z)} + O(\gamma^2)
\ee
of an operator $\O$ with scaling dimension $\Delta$ in $\N=4$ super Yang-Mills
theory actually possesses the Cardy form \cite{Cardy}, \cite{McAO} demanded by
the reduced conformal symmetry.

Let us denote by $C_{\O\O}$ and $C_{\O\O\cl'}$ the coefficients
appearing in the two- and three-point functions
\bea
\aver{\O(x) \O(y)}_{\N=4} &=& \frac{C_{\O\O}}{(x-y)^{2\Delta}},
\nonumber\\
\aver{\O(x) \O(y) \L'(z)}_{\N=4} &=&
\frac{C_{\O\O\L'}}{(x-y)^{2\Delta-4}(x-z)^4(y-z)^4}.
\eea
The renormalized two point function computed via conformal
perturbation theory to order $\gamma$ is given by
\be
\aver{\O(x) \O(y)}_{\rm Janus}^{\rm ren} = Z(x_3)^{-1} Z(y_3)^{-1}
\aver{\O(x) \O(y)}_{\rm Janus} =   C_{\O\O} \, x_3^{-\tilde\Delta(x_3)}
y_3^{-\tilde\Delta(y_3)} G(x,y),
\ee
where $Z(z_3)$ is the renormalization factor
\be
Z(z_3)= 1 + \gamma \pi^2 \l(2\log2\delta+1\r)
\frac{C_{\O\O\L'}}{C_{\O\O}} \, \eps(z_3)
\ee
needed to make to renormalize the composite operator $\O$. The infinities are
regulated via $z_3$ spatial cutoff $\delta$. The corrected anomalous dimension
$\tilde\Delta$ is given by
\be\label{Delta-KK}
\tilde\Delta(z_3) = \Delta + 2\pi^2 \gamma
\frac{C_{\O\O\L'}}{C_{\O\O}} \, \eps(z_3)
\ee
and the nontrivial part of the Green function is
\be
G(x,y) = \l( \frac{x_3 y_3}{(x-y)^2} \r)^\Delta \l[ 1- \gamma \pi^2
\frac{C_{\O\O\L'}}{C_{\O\O}} \biggl( \eps(x_3)+ \eps(y_3) \biggr)
\l(\frac{(x-y)^2}{(x-Ry)^2}  + \log\frac{4(x-y)^2}{(x-Ry)^2} \r)
\r].
\ee
One can easily check that $G(x,y)$ is in fact a function of the dimensionless
ratio $\xi = \frac{(x-y)^2}{4x_3 y_3}$ only as required by the reduced
conformal symmetry, see \cite{Cardy}, \cite{McAO}.

Note that the wave function renormalization factor and the anomalous
dimension are different on both sides of the defect. This could have
been indeed anticipated since from the folding trick perspective\footnote{
See \cite{folding} and references therein.}
one has two independent fields living in one half space only and clearly
those fields are allowed to have different $Z$ factors and
dimensions.

To complete the discussion, let us check now that the dimensions
(\ref{Delta-KK}) exactly reflect the change in the coupling in the
respective half space. Consider therefore a perturbation
\be
\cl_{g^2+\delta g^2}' = \cl' - \frac{\delta g^2}{g^2} \cl'.
\ee
The corrected two-point function can be written as
\be
\aver{\O(x) \O(y)}_{g^2+\delta g^2} = Z^2 \,
\frac{C_{\O\O}}{(x-y)^{2\tilde\Delta}},
\ee
where the corrected anomalous dimension $\tilde\Delta$ is given by
\be\label{Delta-deltag}
\tilde\Delta = \Delta + 2\pi^2 \frac{\delta g^2}{g^2}
\frac{C_{\O\O\cl
'}}{C_{\O\O}}
\ee
which precisely agrees with (\ref{Delta-KK}) upon identification of the
couplings $g^2 \pm \delta g^2 = (1\pm\gamma)g^2 + O\l(\gamma^2\r)$. The $Z$
factors also exactly match with the same identification. Note that the equation
(\ref{Delta-deltag}) is in fact the renormalization group equation
\be
\frac{d\Delta(g)}{dg^2} = \frac{2\pi^2}{g^2}
\frac{C_{\O\O\L'}}{C_{\O\O}}
\ee
valid to all orders in $g$.

\subsection{$\aver{\Tr X^{k_1}\, \Tr X^{k_2}\dots \Tr X^{k_n}}_\g$ in
gauge theory and gravity}

It is quite straightforward to apply the CPT method to $2$-point
correlation functions of the chiral primary operators $\Tr X^k$. The
order $\g$ contribution involves the 3-point function
$\aver{\Tr X^k(x)~\Tr X^k(y)\,\cl'(z)}$ which is also protected. In
the free field limit where we replace $\cl' \to -\frac{1}{4}F^2$,
this correlator vanishes! The chiral primaries are a special case of
Sec. 5.3 in which $C_{\O\O\L'}=0$, and the two-point functions are not corrected
to order $\gamma$.

In Sec 3. we discussed the supergravity result that
$\aver{\Tr X^{k_1}\, \Tr X^{k_2}\dots \Tr X^{k_n}}_\g$ vanishes in the
strong coupling limit for all $n$-point functions. Thus there is a
nice match between gauge theory and gravity for two-point
functions $\aver{\Tr X^k\, \Tr X^k}_\g$, but we do not see how to prove
the more general result in the gauge theory.

\section{An alternate approach to the Janus dual}
\setcounter{equation}{0}

In this section we present an alternate approach to the Janus dual
gauge theory motivated by the D3-brane probe action. This will lead
us to a field theory action suitable for weak coupling perturbation
theory in which propagators for the scalar and spinor fields have
standard form, but the gauge field obeys non-trivial jump
conditions at the defect required by the equations of motion.
Perturbative calculations based on this action are valid to all
orders in $\g$ but fixed order in $g^2_{YM}N$, a perfect complement
to the CPT calculations of Sec.~5.

The (bosonic part of the) action of a single BPS probe D3-brane in a
background solution of Type IIB supergravity in {\it Einstein frame} is
\be \label{d3}
S_{D3} = -T_3 \int d^4\xi \, \sqrt{-\det\l(G_{ab}^E + 2\pi\alpha'
e^{-\frac{\phi}{2}} F_{ ab}\r)} + T_3 \int C_4.
\ee
The key feature is that the dilaton couples only to $F_{ab}$. This
suggests that we study the non-abelian action
\bea \label{newlag}
\cl &=& \Tr\l[-\half D_\m Y^I D^\m Y^I -\half i
\bar{\lam}^\a\slashed{D}\lam^\a - \frac{1}{4g^2(z)} F_{\m\n}^2 \r .\\
\nonumber &+& \l . \frac{g(z)}{2} \bar{\lam}\G_I[Y^I,\lam] +
\frac{g(x)^2}{4}[Y^I,Y^J]^2 \r] \,,
\eea
with $D_\m = \pa_\m +i[A_\m,.]$. This action reduces to $\cn =4$ SYM
theory\footnote{The Yukawa term is given in schematic form as
obtained via dimensional reduction from 10 dimensions.} in the limit
where the position dependent gauge coupling $g(z)$ becomes constant.

The relation between this action and (\ref{janact}) may be seen by scaling
the spinor field by $\lam \to \frac{1}{g(x)} \lam$ and redefining the scalars
by  $Y^I\,=\,\frac{1}{g}X^I.$ The extra term generated in the spinor kinetic
action vanishes for Majorana spinors. For the scalars, after
simple manipulations in which surface terms are dropped, we obtain
\be
- \pa_\m Y\pa^\m Y
                 = \frac{1}{g^2}X \Box X - \frac{\l(\partial g\r)^2}{g^4} X^2.
\ee
For the Janus dual, the last term becomes singular and of order $\g^2$,
so that (\ref{janact}) and (\ref{newlag}) give equivalent results to order
$\g$. Note also that  $\aver{\cl'~ X^2}_{\cn =4}$ vanishes because it involves 
operators
in different multiplets. Thus the order $\g^2$ calculation in Sec. 5 is
not affected by the change in the action.

We propose to use (\ref{newlag}) as the basis for a perturbative approach to
the Janus dual gauge theory. The first step is to define conformal covariant
propagators. For scalars and spinors these are given by the standard
Euclidean propagators
\bea \label{yprop}
\aver{Y^I(x)Y^J(y)}&=& \d^{IJ} \D(x-y)\\
\D(x-y) &=&  {1\over 4\pi^2}\frac{1}{(x-y)^2}\,,
\eea
and
\bea \label{lamprop}
\aver{\lam^\a(x)\lam^\b(y)} &=&\d^{\a\b} \S(x-y)\\
\S(x-y)&=& -\slashed{\pa}\D(x-y)=
 {1\over 2 \pi^2}{\slashed{x} - \slashed{y} \over (x-y)^4}\,.
 \eea

The gauge field action in (\ref{newlag}) is a non-abelian
generalization of the Maxwell action in an inhomogeneous medium with
permittivity $\eps$ and permeability $\mu$ related by $
\eps(z)=\frac{1}{\mu(z)} = \frac{1}{g^2(z)}.$ The speed of light is
constant. In the Janus limit, we have an interface
between two regions with different dielectric constants, a situation
often treated in electrodynamics texts such as \cite{jackson}. The
field strength obeys the following jump conditions obtained by
integrating the equations of motion and the Bianchi identities over
a small pill-shaped region which straddles the interface:
\bea\label{bcf}
\l. F_{ij}(z) \r|_{0^-} &=& \l.  F_{ij}(z) \r|_{0^+} \qquad i,j \ne
3
\\
\nonumber\\
\l. \frac{ F_{i3}(z)}{g^2_-}\r|_{0^-} &=& \l. \frac{
F_{i3}(z)}{g^2_+}\r|_{0^+}.
\eea
The potential obeys
\be\label{bca}
\l. A_i(z)\r|_{0^-} = \l. A_i(z)\r|_{0^+} \l. \qquad \frac{
A_3(z)}{g^2_-}\r|_{0^-} = \l. \frac{ A_3(z)}{g^2_+}\r|_{0^+}\,.
\ee
The propagator may be found by image charge methods. We use the
notation $(Ry)_\m =(y_0,y_1,y_2,-y_3)$ to indicate the image of the
point with coordinates $y_\m$. The reflection matrix is $R_{\m\n} =
{\rm diag}(1,1,1,-1)$. In Feynman gauge the propagator is a
superposition of direct and image terms
\bea \label{arop}
\D_{\m\n}(x-y) &=& {1 \over 4\pi^2 }\frac{\d_{\m\n}}{(x-y)^2}\\
\tilde{\D}_{\m\n}(x,y) &=& {1 \over 4\pi^2
}\frac{R_{\m\n}}{(x-Ry)^2}.
\eea
It is straightforward to obtain the appropriate factors of the
couplings $g_{\pm}$ in the two regions and to write the propagator
as
\bea \label{aprop}
G_{\mu\nu}(x,y) & = & \label{bothpos} \frac{g_+^2}{4 \pi^2 }
\left[\frac{\delta_{\mu\nu}}{(x-y)^2} +  \left( \frac{g_-^2 -
g_+^2}{g_+^2 +
    g_-^2} \right)  \frac{R_{\mu\nu}}{(x-Ry)^2} \right] \theta(x_3)
\theta(y_3) \\
\label{posneg} & + & {1\over 2 \pi^2} {g_+^2 g_-^2 \over (g_+^2 +
g_-^2)}
      {\delta_{\mu\nu}\over (x-y)^2} \Bigl( \theta(x_3)\theta(- y_3) +
      \theta(-x_3) \theta(y_3) \Bigr) \\
\label{bothneg} & + & {g_-^2 \over (4 \pi^2)}
\left[{\delta_{\mu\nu}\over (x-y)^2} +
  \left( \frac{g_+^2 - g_-^2}{g_+^2 + g_-^2}\right) {R_{\mu\nu}\over (x -
  Ry)^2}\right] \theta(-x_3) \theta(-y_3).
\eea

The field strength propagators are superpositions, with the same
coupling factors and step functions as in (\ref{aprop}), of the direct and image terms
\bea \label{fprop}
\<F_{\m\n}(x) F_{\rho\s}(y)\>_{{\rm direct}} &=&\label{fdir}
\frac{1}{\pi^2}\frac{J_{\m\rho}(x-y)J_{\n\s}(x-y) - (\rho
\leftrightarrow \s)}{(x-y)^4}\\
\<F_{\m\n}(x) F_{\rho\s}(y)\>_{{\rm image}} &=&\label{fimg}
\frac{1}{\pi^2}\frac{J_{\m\tau}(x-Ry)J_{\n\lam}(x-Ry) -
(\tau \leftrightarrow \lam)}{(x-Ry)^4} R_{\tau\rho}R_{\lam\s}.
\eea
This propagator is constructed from the conformal Jacobian $J_{\m\n}(z) =
\d_{\m\n} - \frac{2z_\m z_\n}{z^2}$ and transforms correctly under the
conformal inversion.

\subsection{$\aver{\cl'}$ and $\aver{T_{\m\n}}$ again}

We now indicate how to calculate $\aver{\cl'}$ and $\aver{T_{\m\n}}$
in the perturbative formalism and obtain results which agree with
Sec 5. To all orders in $\gamma$ but zero order in $\bar{g}^2$, it
is given by the sum
\be \label{lcalc}
\aver{\cl'(z)} = {\rm Tr}\l[ \half \<\pa_\m Y^I(z)\pa_\m Y^I(z)\> +
\half \<\bar{\lam}^\a (z) \slashed{\pa}\lam^\a(z)\>
+\frac{1}{4\hat{g}^2} \<F_{\m\n}(z) F_{\m\n}(z)\>\r]\,,
\ee
in which the various propagators appear with $y=x$. The divergent
``normal-ordering" terms in (\ref{lcalc}) cancel between the
scalar, spinor, and gluon terms. The same cancellation occurs in
the cosmological constant in a supersymmetric theory. This leaves
the unambiguous finite contribution of $\<F_{\m\n}(x)
F_{\m\n}(x)\>_{{\rm image}}$, which is easily evaluated with the
result
\be \label{lagain}
\aver{\cl'} =  \frac{g_+^2 - g_-^2}{g_+^2 + g_-^2} \frac{3
N^2}{16\pi^2 x_3^4 } \eps(x_3) \, .
\ee
This agrees with (\ref{triumpf}).

Similar reasoning may be applied to the calculation of
$\aver{T_{\m\n}(z)}$. There is a residual unambiguous finite
contribution from the image term of the field strength propagator,
and it vanishes after the index contractions required in $T_{\m\n}$
are done.

\section*{Acknowledgements}
We thank M. Bianchi, E. D'Hoker, A. Hanany, B. Kors, H. Liu, J.
Minahan, A. Nelson, H. Osborn, J. Polchinski, W. Taylor, and D. Tong
for useful discussions and suggestions. The research of DZF is
supported by the National Science Foundation Grant PHY-00-96515.
Both DZF and MS are supported in part by DOE contract
\#DE-FC02-94ER40818. Both ABC and AK are supported in part by DOE
contract \#DE-FG03-96-ER40956.

\begin{appendix}
\section{$\cn =1$ interface SUSY and its failure for $\cn =4$}
\setcounter{equation}{0}

The specific task of this appendix is to show that both proposed actions for
the Janus gauge theory dual explicitly break all supersymmetry. However, we
will proceed toward this goal in an indirect fashion. Namely, we consider
general $\cn =1$ SUSY theories in which the coupling constants $g(z)$ depend on
one of the spatial coordinates, called $z$. We show that two of the four
supercharges can be preserved by adding uniquely determined terms to the
Lagrangian which involve the first derivative $\pa_zg(z)$. In the limit of a
sharp interface, $\pa_zg(z) \to 2 \delta g \d(z)$, so the added terms may be called
``interface counterterms.''

The restriction to two supercharges is achieved by a projection
condition $\Pi \eps = \eps$, where
\be\label{P5z}
\Pi = \frac{1+ i \Gamma^5 \Gamma^z}{2}.
\ee
The preservation of two supercharges is what is expected on general grounds
from the arguments in \cite{dfo}, and the projector (\ref{P5z}) is the same one
that appeared in the defect theory studied in \cite{dfo}. The results for
defect SUSY in a general $\cn = 1$ theory may have useful applications.
However, when we apply them to the $\cn =1$ presentation of $\cn =4$ SYM
theory, we find that the counter terms violate the $SO(6)$ symmetry required
for the Janus dual and are therefore not permitted.

\subsection{N=1 supersymmetry with chiral multiplet}

Let us start with the most general $N=1$ supersymmetric action in
$d=4$ for one chiral multiplet $(\phi, \psi, F)$
\be\label{L chiral}
{\cal L} = - \partial_\mu \phi^* \partial^\mu \phi - \frac{i}{2}
\bar\psi \Gamma^\mu \partial_\mu \psi + F^* F + W' F - \frac{i}{2}
W'' \bar\psi P_+ \psi,
\ee
where $\phi$ is a complex scalar field, $\psi$ is a Majorana fermion and $F$ is
a complex auxiliary field. The superpotential $W$ is a holomorphic function of
$\phi$ and depends on some real couplings $f, m, g, \ldots$, which we shall let
depend arbitrarily on one of the coordinates called $z$. The SUSY
transformation rules are
 \bea
\delta\phi &=& i \sqrt{2} \bar\eps P_+ \psi
\nonumber\\
\delta\phi^* &=& i \sqrt{2} \bar\eps P_- \psi
\nonumber\\
\delta(P_+\psi) &=& \sqrt{2}\l(P_+ F  + \Gamma^\mu
\partial_\mu\phi P_{-} \r) \eps
\nonumber\\
\delta(P_-\psi) &=& \sqrt{2}\l(P_- F^* + \Gamma^\mu
\partial_\mu\phi^* P_{+} \r) \eps
\nonumber\\
\delta(\bar\psi P_-) &=& \sqrt{2} \bar\eps \l(P_- F^* - P_{+}
\Gamma^\mu \partial_\mu\phi^* \r)
\nonumber\\
\delta(\bar\psi P_+) &=& \sqrt{2} \bar\eps \l(P_+ F  -  P_{-}
\Gamma^\mu \partial_\mu\phi \r)
\nonumber\\
\delta F &=& i \sqrt{2} \bar\eps  \Gamma^\mu \partial_\mu P_+ \psi
\nonumber\\
\delta F^* &=& i \sqrt{2} \bar\eps  \Gamma^\mu \partial_\mu P_-
\psi,
\eea
where $P_{\pm}= (1\pm\Gamma^5)/2$ are chiral projection operators.
The variation of the Lagrangian is
\bea\label{delta L}
\delta {\cal L} &=& \frac{i}{\sqrt{2}} \partial_\mu \l[ \bar\eps
P_+ \Gamma^\mu \l(-\Gamma^\nu \partial_\nu \phi^* + F + 2W'^*\r)
\psi + \bar\eps P_- \Gamma^\mu \l(-\Gamma^\nu \partial_\nu \phi +
F^* + 2W' \r) \psi \r]
\nonumber\\
&& -i\sqrt{2}\, \partial_z g \, \bar\eps \l(P_+ \Gamma^z
\frac{\delta {W'}^*}{\delta g} + P_- \Gamma^z \frac{\delta
W'}{\delta g} \r) \psi.
\eea
Here $g$ denotes in fact all possible couplings which depend
nontrivially on the $z$ coordinate. It is important that we have
included all the dependence on $g$ into the superpotential, in
other words that we are using the standard perturbative
normalization for the fields.\footnote{Note that defect SUSY fails for
\be
{\cal L} = -\frac{1}{2g^2(z)}[ \pa_\mu \phi^*\pa^\mu\phi + i \bar{\psi}\G^\m
\pa_\m \psi].
\ee
There are projections $S\eps$ which allow the cancellation of terms
proportional to $\G^\m\pa_\m\phi\psi$ or $\G^\m\pa_\m\phi^*\psi$ in
$\d {\cal L}$, but not both. This remains the case if $ \pa_\mu\phi^*\pa^\mu\phi
\to -\half [\phi^*\Box\phi + \Box \phi^* \phi].$}
This allows us to discard the total
derivative term in (\ref{delta L}). The second term in (\ref{delta
L}) can be simplified by imposing the additional condition on the
supersymmetry parameter $\Pi\eps = \eps$ where
$\Pi$ is given in (\ref{P5z}). The second term in
(\ref{delta L}) then becomes
\be
 + \sqrt{2}\, \partial_z g \, \bar\eps
\l(- P_- \frac{\delta {W'}^*}{\delta g} + P_+ \frac{\delta
W'}{\delta g} \r) \psi,
\ee
which remarkably turns out to be a supersymmetric variation of $ 2
\partial_z g \im \frac{\delta W}{\delta g}$. We can therefore
modify our Lagrangian (\ref{L chiral}) by subtracting this piece
\be \label{chict}
{\cal L}' = {\cal L} - 2 \partial_z g \im \frac{\delta W}{\delta
g},
\ee
such that the new Lagrangian is invariant under the reduced
supersymmetry. Note also that in the limit of constant coupling
$g$ the added piece disappears.

\subsection{N=1 supersymmetry with vector multiplet}

Now let us turn our attention to the $d=4$ nonabelian vector
multiplet $(A_\mu^a, \lambda^a, D^a)$. The most general $N=1$
supersymmetric action without matter fields reads
\be\label{L vector}
{\cal L} = - \frac{1}{4g^2} F_{\mu\nu}^a F^{a \mu\nu} -
\frac{i}{2g^2} \bar\lambda^a \Gamma^\mu D_\mu \lambda^a +
\frac{1}{2g^2} D^a D^a + D^a \xi_a,
\ee
where $A_\mu^a$ is a gauge field, $\lambda^a$ is a Majorana
fermion (gaugino) and $D$ is a real auxiliary field. The constants
$\xi_a$ are Fayet-Iliopoulos terms and are present only for $U(1)$
groups. The coupling $g$ is allowed to depend on one of the
coordinates called $z$.
\bea
\delta A_\mu^a &=& -i \bar\eps \Gamma_\mu \lambda^a
\nonumber\\
\delta\lambda^a &=& \frac{1}{2} \Gamma^{\mu\nu} \eps  F_{\mu\nu}^a
+ i \Gamma^5 \eps D^a
\nonumber\\
\delta\bar\lambda^a &=& -\frac{1}{2} \bar\eps \Gamma^{\mu\nu}
F_{\mu\nu}^a + i \bar\eps \Gamma^5  D^a
\nonumber\\
\delta D^a &=& -\bar\eps \Gamma^5 \Gamma^\mu D_\mu \lambda^a
\eea
The variation of the action (assuming that the FI constants are
indeed constant) is
\be
\delta S = - \int \partial_z \l( \frac{1}{g_a^2} \r) \l[
\frac{i}{4} \bar\eps \Gamma^z \Gamma^{\mu\nu} \lambda^a
F_{\mu\nu}^a + \frac{1}{2} \bar\eps \Gamma^z \Gamma^5 \lambda^a
D^a \r]
\ee
Imposing $P\eps = \eps$ with $P$ given in (\ref{P5z}) we find that
remarkably again the action can be made supersymmetric with the
modification
\be \label{gaugect}
{\cal L}' = {\cal L} + \partial_z \l( \frac{1}{4 g_a^2} \r)
\bar\lambda^a \Gamma^5 \lambda^a.
\ee

The most general $N=1$ supersymmetric theory contains a vector
multiplet and an arbitrary number of
chiral fields in some representation of the gauge group. The
action is given by the sum of (\ref{L vector}) and (\ref{L
chiral}) replacing the ordinary derivatives by the covariant ones
and adding an additional Yukawa term
\be
-\frac{1}{\sqrt{2}} \phi^{i*} t_{ij}^a \bar\lambda^a P_+ \psi^j +
c.c. + \frac{1}{2} D^a \phi^{i*} t_{ij}^a \phi^j.
\ee
Since none of the added terms nor the supersymmetric
transformation rules depend on the non-constant couplings $g_a$,
their supersymmetric variation will add up to a true total
derivative term and can be safely discarded.

\subsection{ $\cn =4$ SYM theory}

In the $\cn =1 $ description of $\cn =4$ SYM theory there are three chiral
multiplets in the adjoint representation with superpotential $W \sim \eps_{ijk}
{\rm Tr}\phi^i\phi^j\phi^k$. One must distinguish between the three fermion
fields $\psi^i$ which are partners of the $\phi^i$, and the fourth fermion
$\lam$, which is the partner of the gluon. One can preserve two supercharges by
imposing the condition $\Pi\eps = \eps$, discussed above and adding the
counterterms in (\ref{chict}) and (\ref{gaugect}). However the largest flavor
symmetry of these counter terms is $SU(3)$ rather that the $SO(6)$ required to
agree with the properties of the bulk Janus solution. Thus the counterterms are
not permitted for Janus, and we conclude that the Janus dual violates SUSY
completely and explicitly.

\section{Some integrals used in Sec. 5.3.}
\setcounter{equation}{0}

This appendix contains two integrals used in the main text. Both of them are 
regulated by imposing a spatial cutoff $|z_3-x_3|>\delta$ and
$|z_3-y_3|>\delta$.
\bea
\int d^4z \, \frac{\eps(z_3)}{(x-z)^4 (y-z)^4} &=& \frac{\pi^2}{(x-y)^4}\l[ -2
\log\delta -1 + \frac{(x-y)^2}{(x-Ry)^2}  + \log\frac{(x-y)^2}{(x-Ry)^2} \r]
\l( \eps(x_3) + \eps(y_3) \r),
\nonumber\\
&&
 + \frac{2\pi^2}{(x-y)^4} \biggl( \log x_3\, \eps(x_3) +\log y_3\, \eps(y_3)
 \biggr)
\\\nonumber\\\nonumber\\
\int d^4z \, \frac{1}{(x-z)^4 (y-z)^4} &=& \frac{2\pi^2}{(x-y)^4}\l[ -2
\log2\delta -1 + \log (x-y)^2 \r].
\eea

\end{appendix}

\end{document}